# Voice-based debate with an AI adversary is associated with increased divergent ideation

Neelam Modi Jain[1,*], Dan J. Wang[1]

[1] Management Division, Columbia Business School, Columbia University, New York, NY 10027
* Correspondence to: njm2180@gsb.columbia.edu

**ABSTRACT**

Concerns that interacting with generative AI homogenizes human cognition are largely based on evidence from text-based interactions, potentially conflating the effects of AI systems with those of written communication. This study examines whether these patterns depend on communication modality rather than on AI itself. Analyzing 957 open-ended debates between university students and a knowledgeable AI adversary, we show that modality corresponds to distinct structural patterns in discourse. Consistent with classic distinctions between orality and literacy, spoken interactions are significantly more verbose and exhibit greater repetition of words and phrases than text-based exchanges. This redundancy, however, is functional: voice users rely on recurrent phrasing to maintain coherence while exploring a wider range of ideas. In contrast, text-based interaction favors concision and refinement but constrains conceptual breadth. These findings suggest that perceived cognitive limitations attributed to generative AI partly reflect the medium through which it is accessed.

## INTRODUCTION

Concerns about generative AI increasingly revolve around its effects on human cognition. A growing empirical literature documents increasing similarity in language and arguments produced during human-AI interaction, particularly in open-ended tasks such as writing and idea generation (1–5). Recent evidence even suggests a cultural feedback loop wherein AI-preferred vocabulary is latently homogenizing the specific words humans use (6). However, this evidence is drawn almost entirely from text-based interfaces, making it difficult to disentangle the effects of AI systems from the communicative medium through which they are used.

Here, we shift attention from the model to the user. Independent of any particular large language model (LLM) or prompt design (e.g., 5), we ask whether the *mode of communication* - talking versus typing - systematically shapes how people generate and organize ideas during a discussion with AI. If communication modality alters the structure of discourse, then effects currently attributed to "AI use" may partly reflect the medium through which cognition is expressed rather than the properties of the AI system itself.

This question has deep intellectual roots. Scholars have long argued that orality and literacy rely on distinct cognitive processes (7). Spoken discourse unfolds in real time and cannot be revised, whereas writing allows for editing and retrospective control (8). As a result, oral exchange tends to be more verbose and more lexically repetitive, while writing supports concision and reduced surface-level redundancy (7). Importantly, repetition in speech is not synonymous with cognitive stagnation: Recurrent phrasing can function as a scaffold that maintains coherence and enables speakers to extend an argument as it unfolds, even as new ideas are introduced (9).

The proliferation of generative AI brings these long-standing distinctions into a new empirical setting. Unlike interactions with human interlocutors, generative AI enables sustained, open-ended exchange while holding constant the content knowledge, argumentative stance, and responsiveness of the opposing party. This allows speaking and writing to be compared as alternative modes of engaging with the same system on an identical task. Open-ended debate is a task that is well-suited for examining these modality effects because it places competing demands on human cognition. Debating requires persistence and repetition to maintain coherence and defend a position and, given that there is no "right answer", it also benefits from the introduction of different arguments and lines of reasoning.

**APPROACH**

The data come from CAiSEY (Classroom Artificial Intelligence Studio for Engaging You), an AI-based discussion system that we designed in-house to support structured, debate-style interaction with an informed AI interlocutor. Study participants were MBA and EMBA students enrolled in an elective course on Technology Strategy at a top U.S. business school. Over the course of the term, students absorbed 16 different case readings, with the option of using CAiSEY to discuss a debate question associated with each of the case readings. This resulted in 957 debate-style CAiSEY conversations for us to analyze from 203 students.

For each case assignment, students were asked to read a case reading that provides necessary background information. Then, if the student chose to engage with CAiSEY, they first selected whether to interact with CAiSEY through a voice-based or text-based interface. When the voice-based option was chosen, the perceived gender of CAiSEY's voice (female versus male) was randomly assigned. To initiate the discussion, CAiSEY prompted the user to take a position on an open-ended strategic question (e.g., "Going forward, what should Netflix prioritize in its content strategy?"). The user then selected a stance (e.g., prioritizing original content over third-party licensing) and provided an opening argument. CAiSEY automatically adopted the opposing position and engaged the user in an extended back-and-forth exchange, responding in real time to the substance of the user's arguments. Aside from communication modality, all interactions were conducted under identical conditions: the same case materials, the same prompts, and the same opposition strategy. Descriptive statistics of the study sample are provided in the SI Appendix.

Because users self-selected their communication modality rather than being experimentally assigned, we cannot make strict causal claims regarding the effect of interface choice on discourse structure. This design introduces two primary sources of potential selection bias: individual-level preferences for a specific format (e.g., due to baseline verbosity or cognitive style) and assignment-level factors (e.g., topic difficulty or path-dependent habituation). Our panel data structure, however, allows us to systematically account for both. To address individual-level selection, our models employ user fixed effects. Descriptive statistics confirm that interactions were not driven by fixed sub-populations: 92% of users (186 of 203) utilized both modalities over the term, and the overall distribution of the 957 interactions was balanced (52% text, 48% voice). To address assignment-level selection, we employ case fixed effects. Furthermore, among successive CAiSEY sessions for the same user (ordered by case sequence), modality differed from the prior CAiSEY session in 52.8% of 754 admissible transitions, demonstrating that users routinely moved between voice and text rather than sticking to a single modality for all cases. This observed switch rate is statistically indistinguishable from a benchmark of independent, equal-probability draws at each such session (two-sided exact test, $p = 0.14$; details provided in the SI Appendix). While this behavioral pattern does not establish experimental randomization, it mitigates concerns of habituation. Consequently, by accounting for these fixed effects and observed switching behaviors, our estimates robustly capture the association between communication modality and discursive output while controlling for confounding characteristics.

**FINDINGS**

Our analysis reveals distinct discursive profiles for spoken versus written interaction ([Fig. 1](#) and [Table 1](#)). First, we find that communication modality is not associated with differences in content originality. Across both conditions, users generated arguments lexically and semantically distinct from the case readings (lexical dissimilarity: $M = 0.64$, $p < 0.001$; semantic dissimilarity: $M = 0.57$, $p < 0.001$), with no significant difference between voice and text (Table 1, Models 1–2).

However, modality is associated with fundamental differences in the structure of the discourse. Consistent with the "copiousness" of oral tradition, voice-based interactions were significantly more verbose, involving more messages (Table 1, Model 3; $\beta = 1.49$, $p < 0.001$) and higher word counts (Model 4; $\beta = 263.4$, $p < 0.001$; Model 5; $\beta = 48.2$, $p < 0.001$) than text-based exchanges. This increased volume was accompanied by greater lexical repetition; talking was associated with a lower type-token ratio (the ratio of unique words to total words, Model 6; $\beta = -0.114$, $p < 0.001$) and, after accounting for conversation length, lower lexical diversity (Model 8; $\beta = -0.042$, $p < 0.05$). These results indicate that spoken discourse relies on a scaffold of recurrent phrasing to maintain coherence.

Crucially, this structural repetition accompanies a broader conceptual search. Despite repeating words more often, voice users exhibited significantly higher divergent ideation - a measure of the range of distinct semantic ideas expressed (Model 9; $\beta = 1.40$, $p < 0.001$). This relationship persists even after accounting for the higher word count of spoken interactions (Model 10; $\beta = 0.178$, $p < 0.001$). Thus, although spoken interactions rely more heavily on repeated words or phrases, they encompass a wider range of ideas.

To illustrate these distinct discursive profiles, consider two representative interactions responding to the same case prompt ("Going forward, what should Netflix prioritize in its content strategy?"). Both users adopted the same initial stance (invest more in third-party licensing) while CAiSEY adopted the opposing stance (invest more in original content). Full transcripts of each interaction are provided in the [SI Appendix](#). In the text-based interaction, the user employed concise language (384 words; TTR = 0.40) to convey three specific arguments in their first message: reduce investment in expensive original content, invest in beloved TV series, and charge a premium fee for box office film releases. In subsequent messages, the user reiterated and refined those initial arguments (e.g., using phrases like "*To clarify, I am not recommending...*") and repeatedly cited the film *The Irishman* as an example of expensive original content, corresponding to a bounded conceptual scope (semantic diversity = 4.51). In contrast, the spoken interaction was nearly twice as long (762 words) and more repetitive (TTR = 0.28), relying on conversational phrasing (e.g., "*I think*", "*you know*", "*when it comes to*"). The initial message introduced broad considerations regarding revenue, subscriber numbers, and pricing constraints. Subsequent messages shifted across a wider array of distinct proposals; alongside suggesting investment in popular TV series, the discussion introduced using AI for content creation, reducing production costs, and pilot testing new consumer experiences. This wider traversal of topics corresponds to a broader conceptual scope (semantic diversity = 5.82).

Finally, exploiting the random assignment of AI voice gender within the spoken condition, we find no evidence that AI persona affects any outcome variable. The modality-related differences persist independent of AI persona characteristics.

**DISCUSSION**

This study suggests that the "homogenization of thought" often attributed to generative AI (4) is tied to the medium through which it is accessed. In an open-ended debate task, talking to an AI adversary is associated with a *greater diversity of ideas* than typing. While spoken interactions are more verbose and lexically repetitive than text exchanges, this redundancy appears to be functional rather than wasteful. By relying on recurrent phrasing to maintain coherence, voice users tend to sustain a broader search of the conceptual space and exhibit higher divergent ideation. One plausible explanation for this result is that voice-based interaction provides paralinguistic cues, such as pauses and intonation (10), that support spontaneous elaboration and lower the cognitive effort required to generate new ideas.

These findings reframe the debate on AI and critical thinking from one of restriction to one of task-interface fit. Rather than treating AI as a monolithic influence on cognition, the results indicate that different interfaces align with different cognitive goals. Crucially, our measure of divergent ideation captures conceptual breadth rather than argument quality. Greater ideational divergence does not imply superior reasoning (5). Rather than privileging one modality over the other, the findings suggest that each is associated with a distinct approach to argument construction. Text-based interaction aligns with exploitation of the argument space - sustaining and elaborating on a focused set of ideas - whereas voice-based interaction aligns with exploration, supporting the generation and articulation of a wider range of arguments in real time. This distinction challenges the assumption that generative AI "outsources" human reasoning (11). Instead, the distinct affordances of typing and talking point to mechanisms through which AI systems can be designed to support, rather than displace, human reasoning.

Several limitations warrant note. The study focuses on MBA and EMBA students engaged in academic debate and does not include a human-only comparison condition, limiting claims about absolute performance relative to non-AI baselines (2). In addition, communication modality was self-selected rather than experimentally assigned, leaving open the possibility that time-varying factors influence modality choice. Future research could randomize modality and examine whether these effects generalize to other task contexts, including collaborative problem solving and creative generation (12).

Plato's Socrates warned that writing would "implant forgetfulness in the soul," yet it also unlocked new forms of reasoning (7). Generative AI may pose a similar challenge: its effects depend not only on what it can do, but on how humans engage with it. This study identifies communication modality as one such point of leverage.

**MATERIALS AND METHODS**

The model results are based on ordinary least-squares regressions with user and case fixed effects and robust standard errors clustered at the user and case levels. All data analysis was conducted using the R statistical software. The SI Appendix contains additional information on measure construction, including the calculation of lexical and semantic diversity. Anonymized data and code to replicate all findings are available for this manuscript.

**ACKNOWLEDGMENTS**

This project was generously supported by the AI in Business Initiative at Columbia Business School. We thank Jill Cohen for helping develop CAiSEY and the MBA and EMBA students at Columbia Business School who opted to use CAiSEY and provided valuable feedback.

**AUTHOR CONTRIBUTIONS**

N.M.J. performed research, analyzed data, and wrote the paper; D.J.W. designed research, performed research, and wrote the paper.

## COMPETING INTERESTS

The authors declare no competing interest.

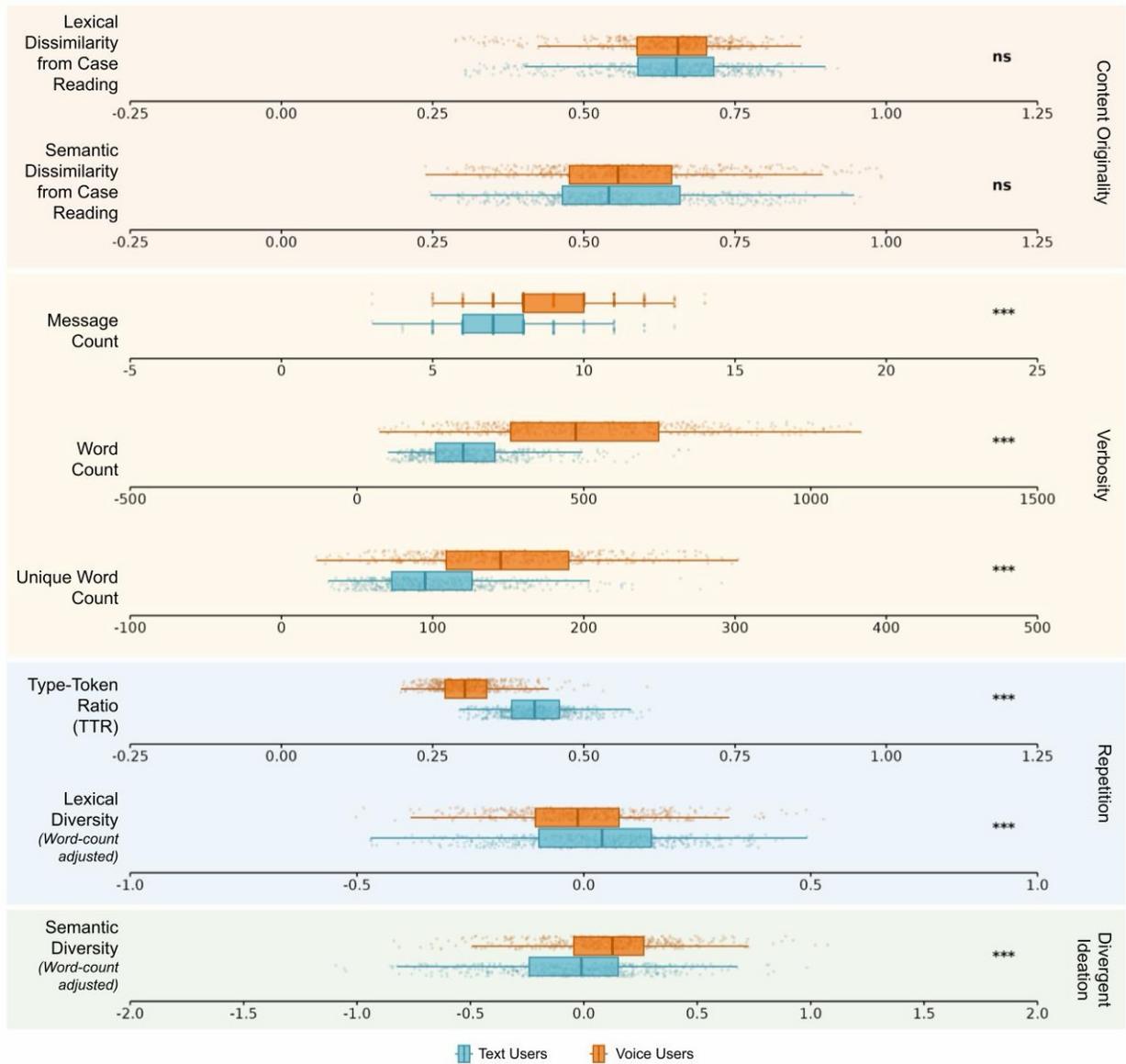

**Figure 1**. **Differences in discourse characteristics between voice- and text-based interaction.** Distributions compare voice-based (N = 458 conversations) and text-based (N = 499 conversations) interaction with the AI adversary. Relative to text-based interaction, voice-based interaction is associated with more messages, higher word counts, greater repetition of words and phrases (lower type-token ratio and lower word-count–adjusted lexical diversity), and higher divergent ideation (word-count–adjusted semantic diversity). Lexical and semantic dissimilarity from the case readings do not differ significantly between modalities, indicating comparable content originality. Points represent individual conversations; boxes show medians and interquartile ranges. Significance indicators denote statistically significant differences in means between voice- and text-based interaction, based on Welch's two-sample t-tests (ns = not significant; + p<0.10, * p<0.05, ** p<0.01, *** p<0.001).

**Table 1.** Results of fixed-effects OLS regressions examining the effect of CAiSEY format on users' conversation characteristics.

| **Dependent Variable:** | **Content Originality** | | **Verbosity** | | | **Repetition** | | | **Divergent Ideation** | |
|---|---|---|---|---|---|---|---|---|---|---|
| | Lexical Dissimilarity from Case | Semantic Dissimilarity from Case | Message Count | Word Count | Unique Word Count | Type-Token Ratio (TTR) | Lexical Diversity | Lexical Diversity (word-count adjusted) | Semantic Diversity | Semantic Diversity (word-count adjusted) |
| | Model 1 | Model 2 | Model 3 | Model 4 | Model 5 | Model 6 | Model 7 | Model 8 | Model 9 | Model 10 |
| CAiSEY Format: Walk-and-Talk | -0.007 | 0.005 | 1.490*** | 263.4*** | 48.2*** | -0.114*** | 0.262*** | -0.042* | 1.4*** | 0.178*** |
| | (0.005) | (0.009) | (0.090) | (12.700) | (3.180) | (0.004) | (0.023) | (0.018) | (0.058) | (0.032) |
| R-squared | 0.812 | 0.755 | 0.518 | 0.697 | 0.655 | 0.766 | 0.647 | 0.101 | 0.739 | 0.186 |
| Within R-squared | 0.004 | 0.000 | 0.232 | 0.501 | 0.327 | 0.649 | 0.215 | 0.014 | 0.530 | 0.076 |

*Notes:* Coefficients represent estimated differences for CAiSEY voice-based users (N = 458) relative to text-based users (N = 499; reference category); All models include user and case fixed effects; Clustered standard errors (by user and case) in parentheses; + p<0.10, * p<0.05, ** p<0.01, *** p<0.001

**SUPPORTING INFORMATION**

**Sample Preprocessing.** The initial dataset included 224 students. We first excluded 17 students who did not use CAiSEY in either the voice or text modalities required for this study. We then excluded an additional 4 students who lacked sufficient data to compute the word-count-adjusted (residualized) lexical and semantic diversity measures used in our fixed-effects regressions. This yielded a final total of 203 students.

**Descriptive Statistics of the Study Sample.** The analytic sample consists of 957 independent CAiSEY conversations generated by 203 MBA and EMBA students across 16 case-based assignments. Of these conversations, 458 were voice-based and 499 were text-based. Within the voice-based interactions, the perceived gender of the AI voice was randomly assigned, with 218 conversations assigned a male voice and 240 assigned a female voice.

Each interaction consisted of a structured back-and-forth exchange between the user and the AI discussion partner. Across all conversations, the total number of messages per interaction averaged 16.47 (SD = 3.40), with a minimum of 6 and a maximum of 28 messages. Because each conversation alternates between user-generated and AI-generated turns, the average number of user-sent messages was approximately half of the total messages, at 7.99 messages per conversation (SD = 1.79), ranging from 3 to 14 messages. Focusing on user-generated messages only, conversations contained an average of 374.64 words per interaction (SD = 218.81), with substantial variation across users and cases (range: 49 to 1,111 words). After removing stopwords, users employed an average of 125.31 unique words per conversation (SD = 53.95), with values ranging from 23 to 302 unique tokens.

**Modality Switching Across Successive CAiSEY Sessions.** Users chose voice versus text for each CAiSEY interaction rather than being randomly assigned. To characterize how often they moved between modalities within user over the term, we defined transitions among successive CAiSEY sessions for the same user. Cases were ordered by the prespecified course sequence. We retained only interactions coded as voice or text and, for each user, compared each CAiSEY session to the immediately preceding CAiSEY session for that user (the first CAiSEY session per user has no predecessor and was excluded). The pooled switch rate is the fraction of all such admissible pairs, pooling across users, in which the two modalities differed. In the analytic sample used here, there were 754 pairs and 398 switches (52.79%).

We contrasted this rate to a simple stochastic benchmark that does not impose any instructional target for how often users should use voice versus text. Conditional on the observed pattern of CAiSEY use (which sessions each user actually completed with CAiSEY), we posited that at each of those sessions modality was drawn independently with probability 0.5 for voice and 0.5 for text. Under that model, the expected pooled switch rate is approximately 50% when the number of admissible pairs is large. We report a two-sided exact binomial test of the hypothesis that the switch probability equals one-half (398 successes in 754 trials; $p$ = 0.135). Because multiple pairs can come from the same user, independence across pairs is only an approximation; we therefore repeated the same data-generating process in a Monte Carlo simulation (10,000 replicates), redrawing i.i.d. modalities at each observed CAiSEY session while preserving each user's count of CAiSEY sessions, which yielded a simulated null distribution centered at 50.01% and an empirical two-sided $p$ of 0.136. This benchmark is intended as a transparency check: it is not a claim of experimental randomization, but it helps assess whether switching behavior is largely consistent with substantial cross-modality use rather than exclusive reliance on a single modality.

**Text Extraction.** User messages were extracted from CAiSEY conversation transcripts by identifying lines prefixed with "USER:" and concatenating them. All text measures were computed on these user-only texts.

**Content Originality:** Lexical dissimilarity between user text and case reading was computed using cosine similarity on document-feature matrices. Both texts were cleaned (removed URLs, email addresses, page numbers, and PDF artifacts), tokenized (removed punctuation, symbols, numbers, and URLs), and converted to document-feature matrices after removing English stopwords (and CAiSEY-specific stopwords: "assistant", "verse", "caisey", "coral" for conversations). Dissimilarity was calculated as $1 - cosine\ similarity$ between the two document vectors. Semantic dissimilarity was computed using sentence embeddings. Both texts were cleaned using identical preprocessing: whitespace normalization, removal of URLs, email addresses, page numbers, dollar amounts, numbers with 5+ digits, all-caps sequences (3+ consecutive words or single words ≥10 characters), case document metadata, and common document structure words (e.g., copyright). Texts shorter than 5 characters after cleaning were excluded. Full cleaned texts were encoded into 384-dimensional vectors using the all-MiniLM-L6-v2 SentenceTransformer model with default parameters (Reimers and Gurevych, 2019[1]). Dissimilarity was calculated as $1 - cosine\ similarity$ between the two embedding vectors.

**Verbosity.** The number of user messages was counted by identifying the lines prefixed with "USER:" in the conversation. Total word count was computed by tokenizing text (lowercased, punctuation removed, split on whitespace), excluding pure numeric tokens, and counting all tokens including stopwords. Unique word count was computed by tokenizing text, removing English stopwords and CAiSEY-specific stopwords, and counting the number of unique tokens remaining.

**Repetition.** Type-token ratio (TTR) was calculated as the ratio of unique words (excluding stopwords) to total words (including stopwords). Lexical diversity was computed as Shannon entropy of the word frequency distribution: $H = -\sum[-p_i * \log p_i]$ where $p_i$ is the relative frequency of word $i$. Because lexical diversity is mechanically related to text length, we used a word-count–adjusted (residualized) lexical diversity measure as the primary outcome in all regression analyses. Specifically, lexical diversity was residualized with respect to word count using a fixed-effects regression with user and case fixed effects. This residualized measure captures variation in repetition net of conversation length. This residualization approach is equivalent to including word count as a control variable in the regression model. We verified this equivalence by estimating models that included both raw lexical diversity and word count; the estimated effect of communication modality was substantively identical across specifications ($\beta$ = -0.105, $p < 0.001$). As expected, raw lexical diversity was strongly correlated with word count ($r = 0.801$), motivating the use of the residualized measure as the main outcome.

**Divergent Ideation.** Semantic diversity was computed as follows. Text was cleaned using the same preprocessing function as for semantic dissimilarity. Texts with fewer than 5 characters or fewer than 3 words after cleaning were excluded. The cleaned text was split into overlapping chunks: starting at word position 0, chunks of 50 words were created with a step size of 25 words (50% overlap), continuing until the end of the text. Chunks with fewer than 3 characters after stripping whitespace were excluded. Texts producing fewer than 2 chunks were excluded. Each chunk was encoded into a 384-dimensional embedding vector using the all-MiniLM-L6-v2 SentenceTransformer model. A pairwise cosine similarity matrix was computed between all chunk embeddings. The diagonal (self-similarities) was set to 0. The upper triangle of the similarity matrix (excluding the diagonal) was extracted, yielding $\frac{n(n-1)}{2}$ similarity values for $n$ chunks. Similarity values were clipped to the range [0, 1] to ensure non-negativity. If the sum of similarities was zero, the value was set to missing. Similarities were normalized to probabilities by dividing each value by the sum of all similarities. Zero probabilities were removed. Shannon entropy was computed on the resulting probability distribution: $H = -\sum[-p_j * \log p_j]$ where $p_j$ is the normalized similarity value for pair $j$. As with lexical diversity, semantic diversity was highly correlated with word count ($r$ = 0.943). Because this correlation was substantially stronger than in the lexical case, including both semantic diversity and word count in the same regression led to severe multicollinearity. Accordingly, we used a word-count-adjusted (residualized) semantic diversity measure as the primary outcome. Semantic diversity was residualized with respect to word count using a fixed-effects regression with user and case fixed effects. This approach isolates variation in ideational breadth net of conversation length and is substantively equivalent to controlling for word count, without introducing collinearity into the regression model.

---

[1] N. Reimers, I. Gurevych, Sentence-bert: Sentence embeddings using siamese bert-networks. arXiv preprint arXiv:1908.10084 (2019).

**Statistical Analyses.** All regressions were implemented in the R statistical software. They are ordinary least-squares regressions with user and case fixed effects and robust standard errors clustered at the user and case levels.

**Try CAiSEY Yourself:** A live demonstration of the voice-based CAiSEY interface is available at https://caisey.me/demo.

**Representative Transcripts.**
**Case Subject:** Netflix
**Debate Prompt:** Going forward, what should Netflix prioritize in its content strategy?
- **Stance 1:** Invest more in original content than third-party licensing
- **Stance 2:** Invest more in third-party licensing than original content

In both interaction transcripts below, note that the users chose Stance 2 thus CAiSEY adopted Stance 1.

## Text-Based Interaction

**USER105:** They should spend less on original content moving forward, but should still produce similar quantities -- instead focusing on bingeable or indie content with lesser known/rising talent. They should not be investing in expensive other projects like Irishman, especially if the content does not align with their customer's preferences. Separately, Netflix should continue to invest in third-party content for new film releases (with a premium fee) and also beloved TV series that customers return to again and again.
--
ASSISTANT: That's an interesting position. On a scale of 1 to 5, 1 being not very confident and 5 being very confident, how confident do you feel about your argument?

--
**USER105:** 4

--

ASSISTANT: Great, thanks for sharing your confidence level. I am going to take an alternate position for the sake of this discussion. I will argue that Netflix should prioritize original content over third-party content. Feel free to ask any clarification questions if you do not understand my points. My first challenge to your argument is that your strategy seems to rely heavily on third-party content. This may make Netflix more dependent on other content providers, which is a risky position considering the current trend of media companies pulling their content to start their own streaming services. How would you respond to this?

--

**USER105:** I agree that that is a concern, but, especially as the generation ages, many users will demand their favorite shows (Friends, Golden Girls, ER, Seinfeld, West Wing, etc.). If Netflix does not offer these staples, they could permanently lose customers. Especially since the other competitors that will offer these beloved shows will also offer their own original content.

--

ASSISTANT: While attracting and retaining customers with third-party content may work in the short term, in the long run, it's a question of survival in an increasingly competitive market. By investing in original content, Netflix can differentiate itself and maintain control over its programming. For example, HBO has successfully built its brand around high-quality original programming like "Game of Thrones". Isn't it possible that shifting focus to original content can help Netflix stand its ground in the face of competition?

--

**USER105:** HBO is a leader in high-quality content production, which works well with their model producing fewer projects and releasing episodes over time. Netflix has a very different brand and strategy. They produce mass quantities and sees what sticks. It would be very disruptive (and not in a good way) for Netflix to try to change this perception -- they could alienate customers. To clarify, I do think Netflix should still create a lot of original content. I just believe they should spend much less per project (by focusing on rising talent and creators from non-traditional film regions/cities) and focus on quantity.

--

ASSISTANT: Your argument about sticking to their current model makes sense, but consider Amazon Prime Video's strategy. They balance both original and third-party content while also investing in high-budget projects such as "The Lord of the Rings" series. They have managed to carve a niche for themselves in the crowded streaming market. Netflix could potentially benefit from a mixed strategy that involves creating high-budget original content that can act as a flagship for the service while also investing in third-party content. What's your take on this?

--

**USER105:** To clarify, I am not recommending that they stick to their current model. I want them to pivot away from sports partnerships and very expensive film productions since that is a distraction from their customer base. I also recommended a new suggestion that they charge a premium for new film releases (attained through 3rd party licenses) like Amazon Prime. That said, there may be a benefit for Netflix investing in 1-3 major, flashy, high-quality TV series productions (NOT a film like Irishman) a year, but I would recommend that they make the most of those and release episodes over time (like HBO, Amazon Prime, and Apple TV do).

--

ASSISTANT: Would you like to continue this discussion, or are you ready to wrap up?
--
**USER105:** I'd like to continue
--
ASSISTANT: Alright, let's proceed. You've suggested a hybrid model of content strategy for Netflix, including incremental changes to their current model and a premium fee for new releases. But how about Netflix's global expansion strategy? Original content, especially local-language productions, have proven successful for Netflix in international markets. By prioritizing original content, Netflix can cater to local tastes and preferences, facilitating its global expansion. How would you respond to this argument?
--
**USER105:** Yes, this was one of my initial suggestions.
--
ASSISTANT: Apologies for misunderstanding your position. So, you are indeed in favor of original content, but with a focus on local-language productions and emerging talent, which could be less costly. Considering this, do you think Netflix should perhaps prioritize original content that is strategically aligned with its global expansion and customer preferences, rather than focus on expensive high-profile projects?
--
**USER105:** Yes that is precisely what I recommend! Apologies if I was not clear.
--
ASSISTANT: Great, we are on the same page, then. Shall we continue the conversation, or would you like to wrap up?
--
**USER105:** We can wrap up
--
ASSISTANT: Alright, let's summarize. You argued that Netflix should continue to invest in third-party content and produce lower-budget original content by focusing on rising talent and creators from non-traditional film regions/cities. You also suggested a premium fee for new releases. I argued that Netflix should prioritize original content that is strategically aligned with its global expansion and customer preferences, and should consider investing in a few high-profile projects to serve as flagships.
Now, do you think I have convinced you to agree with my position?
--
**USER105:** You took my suggestion hah

### Voice-Based Interaction

**USER015:** So I would say that Netflix should prioritize third-party content and for the following reasons. I think they've been pursuing the prioritizing original content for quite a while now and we've seen that subscriber growth has certainly slowed down by 2023, meaning that there is something to be said about a different avenue that provides a fresh kind of avenue for Netflix's customer base to really enjoy. We also see that you know when it comes to revenue there really is kind of two components which is a numbers description and a prices description. When it comes to these prices of the customers of Netflix they have already enacted a couple of price increases which have you know again like I think without having additional value it would be hard to justify another price raise. Therefore I believe that adding third-party content will allow Netflix to diversify its offerings, allow them to justify the new price raise and be able to capture a new segment of customers.
--
ASSISTANT: On a scale of 1 to 5, how confident do you feel about your position? 1 being not very confident and 5 being very confident.
--
USER015: I feel like I'm on a scale of 4 in terms of confidence here.
--
ASSISTANT: Got it. I'll take the position that Netflix should prioritize original content. Focusing on original content can differentiate Netflix in a crowded market and build a unique brand identity. This strategy can also lead to long-term cost savings by reducing dependency on expensive licensing deals. For example, shows like "Stranger Things" have become cultural phenomena, driving subscriber retention and attracting new viewers. How do you respond to this argument?
--
USER015: Well I think that certainly there is original content that does generate that but I wonder if the market has become saturated with original content especially given the rise of other competitors like HBO, Hulu, Amazon and Disney Plus and so I wonder if nowadays when it comes to the original content market it's become saturated to the point where now it's looking at these third-party licensing content like with boxy matches or old TV shows that invoke nostalgia that allow for these competitors to really diversify their offerings to the customer base.
--
ASSISTANT: You raise a valid point about market saturation. However, I believe Netflix's scale gives it an advantage. With a content budget exceeding $17 billion, Netflix can outspend rivals on original programming, producing a high volume of successful series and films. Moreover, exclusive original content can create a must-have perception that attracts and retains subscribers. For instance, "The Crown" and "Bridgerton" have both driven significant viewership. What's your response to the notion that scale in original content can differentiate Netflix and drive business performance?
--
USER015: I think that's a very valid point, and Netflix certainly doesn't need to de-prioritize original content. If anything, I feel like they can actually release some of resources to focus on third-party content, for example, by implementing things like AI when it comes to creating original content, allowing them to have, you know, higher productivity when it comes to creating original content, allowing Netflix to focus kind of on other things that are potentially a lot more worth squeezing out.
--
ASSISTANT: Interesting argument. However, I'd challenge the notion that AI can fully replace the creative process in generating original content. Creativity is nuanced and subjective, often requiring a human touch. Moreover, if Netflix diverts resources from original content, it could affect the quality and uniqueness of its offerings, potentially risking its competitive edge. For instance, if Netflix only uses AI for content creation, it might struggle to produce the kind of emotionally resonant shows that have historically driven its success. What's your response to this perspective?
--

**USER015:** I don't think Netflix necessarily needs to fully use AI when it comes to creating the entire lifespan of original content. I actually think that there's a partial solution or even a pilot solution that exists here. So partially, what I'm talking about is that, you know, you don't need to use AI for the entire creation of original content. You can use it to speed up certain processes that are more manual. For example, like, you know, when it comes to potentially like drying out certain kind of scenes when it comes to producing original content or helping editors find grammatical errors or things like that. You know, I think there's really certain parts you can automate and make much faster. Additionally, I think when it comes to if you do want to go down the route of using AI to produce entire original content, you can also pilot it and just see their reactions. I do think that, you know, it's hard to say whether or not consumers will enjoy or not enjoy a show that's been fully AI produced. So you can really pilot this to see how that works.
--
ASSISTANT: You make a fair point about the potential to automate certain aspects of content creation. However, relying on AI, even partially, could lead to homogenization of content. The distinctiveness of original programming is often what attracts and retains subscribers. If content becomes formulaic, it may fail to capture the audience's imagination. Moreover, piloting AI-generated content could be a risky strategy, potentially diluting the brand if the audience rejects it. Consider Coca-Cola's New Coke debacle in the '80s; experimentation can carry significant brand risks. What's your take on this risk factor?
--
**USER015:** Well, I think for one, you know, if you look at Coca-Cola as well, they recently had a very successful marketing scheme with their Dahle and OpenAI partnership. So I don't think that necessarily, you know, when you think about AI, it dilutes a brand as well as the fact that when I think about AI as a whole and how it can be deployed in creativity, I don't think you necessarily need to use AI to be extremely formulaic. Of course, you can have a human overseeing this. It's really just more about having AI help speed up certain processes in the creation process that may be highly manual and not impactful at all to the final creative output.
--
ASSISTANT: Fair enough. Leveraging AI as a tool rather than a replacement could indeed optimize the production process without compromising creativity. Coca-Cola's recent success with AI-driven marketing suggests potential for AI to add value in creative fields. That said, the key challenge for Netflix will be to strike the right balance between automation and preserving the unique storytelling that has defined its brand. Would you like to continue discussing this, or are you ready to wrap up?
--
**USER015:** I am ready to wrap up.
--
ASSISTANT: Great discussion. You argued for Netflix prioritizing third-party content, highlighting potential market saturation and suggesting efficiency improvements with AI. I argued for the focus on original content, emphasizing differentiation and brand identity. Have I convinced you to lean towards prioritizing original content?
--
**USER015:** I think certainly, yes, I originally thought that the route was to go fully on third-party but I do think there is definitely still a lot more to squeeze out of original content. So I do think that I am leaning more towards a more balanced approach here.